\documentclass[twocolumn=true,paper=letter,font size=10,abstract=true,bibliography=openstyle]{scrartcl} 

\usepackage[english]{babel}
\usepackage[ansinew]{inputenc}
\usepackage{textcomp}
\usepackage{graphicx}  
\usepackage{flafter}  
\usepackage{dsfont}
\usepackage{amsmath,amssymb}  
\usepackage{bm}  
\usepackage{pdfsync}

\usepackage[total={18.2cm,25cm},top=2cm,centering, includefoot]{geometry}
\title{Ring sequence decomposition of an accretion disk:\\ the viscoresistive approach}

\author{R. Benini$^{1}$ \and G. Montani$^{1,2,3}$ \and J. Petitta$^{1}$}
\publishers{ { $^{1}$  Department of Physics  - ``Sapienza'' Universit\`a di Roma, \\
Piazzale A. Moro, 5 (00185), Rome, Italy\\
 $^{2}$ ENEA - C.R. Frascati (Department UTFUS-MAG)  \\
   Via Enrico Fermi, 45 (00044), Frascati (Rome), Italy\\
  $^{3}$ Istituto Nazionale di Fisica Nucleare (INFN) - Sezione di Roma 1
} }

\begin{document}

\maketitle

\newcommand{\reff}[1]{(\ref{#1})}
\newcommand{\eref}[1]{eq.\reff{#1}}
\newcommand{\erefs}[1]{eqs.\reff{#1}}
\newcommand{\Eref}[1]{Equation \reff{#1}}
\newcommand{\Erefs}[1]{Equations \reff{#1}}
\newcommand{\tref}[1]{table \reff{#1}}
\newcommand{\fref}[1]{fig.\reff{#1}}


\newcommand{\Div}[1]{\vec{\nabla} \cdot \vec{#1}} 
\newcommand{\grad}{\vec{\nabla} \!} 
\newcommand{\Rot}[1]{\grad \times \vec{#1}}
\newcommand{\partialdiff}[2]{\frac{\partial #1}{\partial #2}} 
\newcommand{\totaldiff}[2]{\frac{\mathrm{d} #1}{\mathrm{d} #2}} 
\newcommand{\abs}[1]{\left| #1 \right|} 
\newcommand{\de}[1]{ \, \mathrm{d} #1} 
\newcommand{\<}[1]{\left(#1\right)} 


\newcommand{\psiD}{\psi_{\mathrm{D}}}
\newcommand{\psiDo}{\psi_{\mathrm{D0}}}
\newcommand{\Mstar}{M_{\mathrm{*}}}
\newcommand{\Md}{M_{\mathrm{d}}}
\newcommand{\Msun}{M_{\mathrm{\odot}}}
\newcommand{\dMd}{\dot{M}_{\mathrm{d}}}
\newcommand{\muo}{\mu_{0}}
\newcommand{\vB}{\vec{B}}
\newcommand{\vJ}{\vec{J}}
\newcommand{\omegaK}{\omega_{\mathrm{K}}}
\newcommand{\omegaKO}{\omega_{\mathrm{K0}}}
\newcommand{\omegaO}{\omega_{\mathrm{0}}}
\newcommand{\omegaP}{\omega^{\prime}}
\newcommand{\omegaPO}{\omega^{\prime }_0}

\newcommand{\vnabla}{\vec{\nabla}}
\newcommand{\DeltaE}{\Delta_\epsilon}

\newcommand{\mD}[1]{{\mathcal{D}}[#1]}
\newcommand{\visco}{\mathds{D}}
\newcommand{\vs}{v_{\mathrm{s}}}
\newcommand{\vt}{v_{\mathrm{t}}}
\newcommand{\vA}{v_{\mathrm{A}}}
\newcommand{\me}{m_{\mathrm{e}}}
\newcommand{\mi}{m_{\mathrm{i}}}
\renewcommand{\ne}{n_{\mathrm{e}}}

\newcommand{\Rstar}{R_{\mathrm{*}}}
\newcommand{\Ls}{L_{\mathrm{s}}}
\newcommand{\Kb}{K_{\mathrm{B}}}

\newcommand{\DER}[3]{\displaystyle\frac{\partial^{#3} #1}{\partial #2^{#3}}}
\newcommand{\DERT}[3]{\displaystyle\frac{d^{#3} #1}{d #2^{#3}}}
\newcommand{\DERO}[3]{\displaystyle\frac{\partial #1}{\partial #2^{#3}}}

\begin{abstract}
We analyze a two dimensional viscoresistive magneto-hydrodynamical (MHD) model for a thin accretion disk which reconciles the crystalline structure outlined in \cite{C05,CR06} with real microscopic and macroscopic features of astrophysical accreting systems. In particular, we consider small dissipative effects (viscosity and resistivity, characterized by a magnetic Prandtl number of order unity), poloidal matter fluxes and a toroidal component of the magnetic field. These new ingredients allow us to set up the full equilibrium profile including the azimuthal component of the momentum conservation equation and the electron force balance relation.  These two additional equations, which were identically satisfied in the original model, permit us to deal with non zero radial and vertical matter fluxes, and the solution we construct for the global equilibrium system provides a full description of the radial and vertical dependence within the plasma disk. The main issue of our analysis is outlining a modulation of the matter distribution in the disk which corresponds to the formation of a ring-like sequence, here associated with a corresponding radial oscillation of the matter flux.
\end{abstract}

\section{Introduction}

The relevance of the accretion processes in understanding the emission properties of many astrophysical sources like Gamma Ray Bursts  \cite{ReviewPiran} and Active Galactic Nuclei \cite{LibroAGN}, led to a significant effort over the last four decades in order to construct a standard model for the profile of an accreting disk over a compact object.  Starting from the original idea of Shakura \cite{S73} a wide number of results (see for istance \cite{pringlerees72}, \cite{Shakura:1988p104}, or \cite{lyndenpringle74,L96}) allowed to achieve the firm understanding that, within a disk,  the angular momentum transport is driven by dissipative effects (for a detailed review see \cite{Bisno01}). Since the nature of such viscosity and resistivity of the plasma cannot be recovered on a microscopical level (such systems are mainly collinsionless), then the origin of non-ideal features must rely on the turbulent behaviour of the plasma once linear instabilities are triggered. The most successful paradigm to explain the onset of the turbulence is the so called Magneto-Rotational Instability, discovered by Velikhov \cite{V59} and Chandrasekhar \cite{1960PNAS...46..253C}, and recently fully developed by Balbus and collaborators \cite{1991ApJ...376..214B} (for a review see also \cite{B98}). However, such instability, which can be triggered by an arbitrary small magnetic field, is nonetheless suppressed when the sound speed is smaller than the Alfv\'en velocity, i.e. when the $\beta$ parameter of the plasma is smaller than one. Since this condition can take place in sufficiently cold and magnetized plasma disks, a challenging puzzle arises  when dealing with the accretion mechanism in structures exhibiting such regimes. 

Two interesting proposals for alternative scenarios were proposed in \cite{2000ApJ...529..978B} and \cite{Coppi:2001p102}. In particular, B. Coppi and coworkers developed over the last ten years a line of research devoted to construct a different paradigm \cite{CoppiHighEnergy}, mainly based on the implementation on the astrophysical setting of some basic features observed in laboratory plasma, like ballooning mode instabilities \cite{Coppi:2003p112}. More than solving the problem of accretion for an ideal plasma disk, such efforts had the main merit to outline the existence of regimes where the disk profile can fragment into a sequence of rings. This situation takes place as soon as the Lorentz force, produced by the plasma backreaction in the magnetic field of the central object, plays a significant role in the vertical confinement. This result, however, was obtained in a local description of a purely rotating disk, and disregarding any dissipative effect (see also \cite{2010PhRvE..82b5402M} for the jet formation in this scheme). Despite being rather small, such dissipation features are present in any real astrophysical system, and the possibility to reconcile the ring sequence with a realistic equilibrium configuration is the main task addressed by the present work.

We consider the full MHD equilibrium including, with respect to the original model, the azimuthal momentum conservation equation and the electron force balance relation. For a purely ideal and rotating plasma, these equations are identically satisfied, but in a real system (here we consider non zero viscosity and resistivity coefficient, associated to a magnetic Prandtl number of order unity) they are crucial for fixing the poloidal matter fluxes and the toroidal magnetic field. We provide a solution for the global configuration system which has the merit to demonstrate that the existence of the ring sequence within the disk is compatible with non zero dissipative effects, and, over all, with the presence of a non zero radial matter flux. Nonetheless, the smallness of the additional terms we include in our treatment, allows us to neglect the poloidal velocity and the toroidal magnetic field in the radial and vertical configuration, which therefore retain the same form as in \cite{C05,CR06}. We emphasize that dealing with only small corrections is a constraint imposed by the corotation theorem \cite{F37}. In fact, as shown in the last section, up to the leading order, it is still possible to preserve the relation between the angular velocity of the disk $\omega$ and the magnetic surface function $\psi$ as far as the additional contributions due to the poloidal velocity, the toroidal magnetic field and the dissipative effects, are sufficiently small with respect to the basic model quantities as in \cite{C05,CR06}. A significant improvement that our analysis is able to provide, is fixing the vertical dependence of the disk profile as a consequence of a compatibility equation coming out from the azimuthal and the electron force balance system. The local model we provide can be regarded as the starting point for the study of the settlement of linear instabilities in the paradigm of the crystalline structure. 

As outlined in \cite{2011PhRvE..84b6406M}, the crystalline morphology of the magnetic field, as well as the ring sequence profile, have a characteristic scale which turns out to be microscopical on an astrophysical setting (indeed the typical length of the perturbations results to be of the order of some meters). This issue suggests that, more than for its stationary appearance, such a morphology concerns the stability structure of the disk, and only after a full scheme of perturbation analysis is developed, a precise nature of the configuration can be settled down. In this respect, our treatment stands as the natural and realistic framework where to study the stability of the plasma disk.

\section{Two-Dimensional MHD Model for an Accretion Disk \label{sec:4}}

We will now set up the basic 2D MHD configuration equations for an axially-symmetric disk around  a compact (few  Solar masses) and strongly magnetized (a dipole-like field of about $10^{12}$ gauss) object, i.e. a typical pulsar source. In such a system, we neglect the self-gravity of the disk, so that the gravitational potential $\chi$, in cylindrical coordinates $\<{r,\,z,\,\phi}$, has the form
\begin{equation}
\chi (r\, ,z) = \frac{G\Mstar}{\sqrt{ r^2 + z^2}}
\, ,
\label{Gravpot}
\end{equation}
where $\Mstar$ is the mass of the central body, and $G$ the gravitational constant.
It is worth noting that
the axial symmetry prevents any dependence
on the azimuthal angle $\phi$ of all the quantities involved in
the problem. The continuity equation associated to the mass conservation, i.e.
\begin{equation}\label{conteq}
\vec{\nabla} \cdot \<{\epsilon \vec{v}} = 0\,,
\end{equation}
admits the following explicit solution
\begin{equation}
\epsilon \vec{v} =
\frac{1}{r}\vec\nabla\Theta\wedge \hat{e}_{\phi } + \epsilon \omega r \hat{e}_\phi
\, .
\label{solconteq}
\end{equation}
Here $\epsilon$ denotes the mass density, $\vec{v}$ the velocity field of the disk, and $\Theta (r\, , z)$ is a function which is odd with respect to the $z$ variable, in order to deal with a non-zero accretion rate $\dot{M}_\mathrm{d}$. In fact, if $H$ is the half-depth of the disk,  $\dot{M}_\mathrm{d}$ results to be given by \cite{Bisno01,2010GReGr.tmp..112M}
\begin{equation}
\dot{M}_\mathrm{d} = -2\pi r\int _{-H}^{H}\epsilon v_rdz =
4\pi \Theta (r\, , H) > 0
\, .
\label{Mdot}
\end{equation}
The dynamics of such a stationary, viscoresistive system is then described by the following equation \cite{Bisno01}
\begin{equation}\label{navstokes}
\begin{split}
\epsilon\<{\vec{v}\cdot\grad}\vec{v} =- \grad\<{p + \displaystyle\frac{B^2}{8\pi}} + \displaystyle\frac{1}{4\pi}\<{\vec{B}\cdot\grad}\vec{B} +\\
+ \epsilon \grad \chi +\nabla_k\left[\visco\<{\nabla_k\vec{v} + \nabla\vec{v}_k - \displaystyle\frac{2}{3} \<{\grad\cdot\vec{v}}} \mathds{I}\right]\,,
\end{split}
\end{equation}  
 $\visco$ being the viscosity and  $\mathds{I}$ the unit tensor; the magnetic field $\vec{B}$, because of the axial symmetry, can be expressed through the magnetic flux function $\psi=\psi\<{r\, ,z^2}$ in the general form 
\begin{equation}
\vB = \frac{1}{r}\vec\nabla\psi\wedge\hat{e}_\phi + 
\frac{I}{r}\hat{e}_{\phi } \, ,
\label{vectorb}
\end{equation}
where $I = I(\psi \, , z)$ is a generic function of $\psi$ characterizing the toroidal component of the field. 
The similarity of the magnetic field and
matter flux structure, is due to their common divergence-less nature.
A further assumption in our derivation is that the angular frequency of the disk $\omega$ has to be a function of  $\psi$, i.e.
\begin{equation}\label{omegadipsi}
\omega=\omega\<{\psi}\,.
\end{equation}
This result has been firstly demonstrated by Ferraro in \cite{F37} under the hypotheses of an ideal, non resistive, axisymmetric rotating plasma. It can be shown (see the detailed proof discussed in the last section) that this result holds even in the presence of small resistivity and viscosity.

\subsection{Local perturbative analysis}

In the present analysis, we are interested in the behavior of the configuration around a fixed, fiducial value $r_0$ of the radial variable $r$. In order to investigate the effects induced on
the disk profile by the electromagnetic reaction of the plasma, we split the mass density and the pressure
contributions in the zeroth (barred) and the first (hatted) orders components, i.e.
\begin{equation}
\epsilon = \bar{\epsilon }(r_0,\, z^2) + \hat{\epsilon}\,,\quad
p = \bar{p}(r_0,\, z^2) + \hat{p}\,,
\end{equation}
respectively. 
The same way, we express the magnetic surface function as the sum of a term $\psi_0$, describing the field of the star, plus a correction $\psi_1$
\begin{equation}
\psi = \psi _0(r_0) + \psi _1(r_0\, , r-r_0\, ,z^2)\,,\qquad \psi _1\ll \psi _0\,.
\end{equation}
The  quantities $\hat{\epsilon}$, $\hat{p}$ and $\psi _1$ describe the change induced by the currents which rise
within the disk. In general, these corrections are
small in amplitude but with a very short scale of variation.
Having this scenario in mind, we address the so-called ''drift ordering'' for
the behavior of the gradient amplitude, i.e.
the first order gradients of the perturbations are of
zeroth-order, while the second order ones dominate.
Accordingly, the profile of the toroidal
currents rising in the disk, has the form
\begin{equation}
J_{\phi } = \displaystyle\frac{c}{4\pi} \vec{\nabla}\wedge \vec{B}\simeq \frac{-c}{4\pi r_0}
\left(\partial ^2_r\psi _1 + \partial_z^2\psi _1\right) = \frac{-c}{4\pi r_0}\Delta\psi_1
\, .
\label{jphi}
\end{equation}
On the other hand, the azimuthal component of the
Lorentz force is related to the existence of the
function $I(\psi ,z)$ and  results to be equal to
\begin{equation}
F_{\phi } \simeq \frac{1}{4\pi r_0^2}
\left(\partial _zI\partial _r\psi -
\partial_rI\partial _z\psi\right)
\, .
\label{fphi}
\end{equation}

As a consequence of the corotation theorem, in the present split scheme, 
we can  decompose  $\omega$ in the sum of the Keplerian term  $\omegaK$  and a constant $\omegaPO$, respectively given by
\begin{equation}\label{splitomega}
\omega = \omega_\mathrm{K} +
\omegaPO\psi_1\,
\end{equation}
\begin{equation}\label{omegak}
\omegaK \equiv \omega\<{\psi_0} = \displaystyle\sqrt{\frac{G\Mstar}{r_0^3}}\,,\quad\quad \omegaPO\equiv\left.\DERT{\omega}{\psi}{}\right|_{\psi_0}
\end{equation}  
This form for $\omega$
holds locally, as far as $(r - r_0)$ remains a sufficiently
small quantity, so that the dominant deviation 
from the Keplerian contribution is due to $\psi_1$ only.

\subsection{Vertical dynamics}

As developed in \cite{C05,CR06}, we can distinguish
the fluid components from the 
electromagnetic back-reaction. The vertical component of \eref{navstokes} implies that, at the zeroth order, the background profile is determined by the pure thermostatic equilibrium
standing in the disk when the vertical gravity
(i.e. the Keplerian rotation) is sufficiently high
to provide a confined thin configuration
\begin{equation}
\begin{split}
D(z^2) \equiv \frac{\bar{\epsilon}}{\epsilon _0(r_0)} = \exp\left(-\frac{z^2}{H^2}\right)\\
\epsilon _0(r_0) \equiv \epsilon (r_0,\, 0)
\, , \hspace{5mm} H^2 \equiv \frac{4\Kb\bar{T}}{m\omegaK^2}\;,
\end{split}
\end{equation}
Here we have assumed that the background obeys an isothermal relation, and that the temperature $T$ may be then represented as
\begin{equation}
2 \Kb T \equiv m\frac{p}{\epsilon} =
m\frac{\bar{p} + \hat{p}}{\bar{\epsilon} +
\hat{\epsilon}} \equiv 2 \Kb (\bar{T} + \hat{T})
\, ,
\label{temptot}
\end{equation}
$\Kb$ being the Boltzmann constant and $m$ the ion mass.
On the other hand, at the first order, the vertical equilibrium is described by the following equation
 \begin{equation}\label{verticalequilibrium}
\partial _z\hat{p} + \omega ^2_\mathrm{K}z\hat{\epsilon}
- \frac{1}{4\pi r_0^2}\Delta \psi_1
\partial _z\psi_1 = 0 
\, .
\end{equation}

\subsection{Radial equilibrium}

The radial component of \eref{navstokes} fixes the equilibrium features
of the rotating layers of the disk, and can be decomposed
into the dominant character of the Keplerian angular
velocity, plus an equation describing the
behavior of the deviation, i.e.
\begin{equation}
\begin{split}\label{radialequilibrium}
2\omega _\mathrm{K}r_0(\bar{\epsilon} + \hat{\epsilon})
\omega _0^{\prime }\psi _1 +
\frac{1}{4\pi r_0^2}\Delta\psi_1
\partial _r\psi_0 =\\
=\partial _r\left[
\hat{p} + \frac{1}{8\pi r_0^2}
\left(\partial_r\psi_1\right)^2\right]
+ \frac{1}{4\pi r_0^2}\partial_r\psi _1 \partial^2_z\psi_1\,.
\end{split}
\end{equation}
Here we neglected  the presence of the poloidal currents,
associated with the azimuthal component of the magnetic
field, as soon as their contribution results to be of higher order.

\subsection{Azimuthal equation}\label{sec:azimuthaleq}

By inserting \eref{splitomega} into the $\phi$-component of \eref{navstokes}, we easily obtain the relation governing the azimuthal equilibrium; more precisely we get
\begin{equation}
\begin{split}
\label{exazeq2}
&\epsilon rv_r\partial _r\psi +
\epsilon rv_z\partial _z\psi +
2\epsilon v_r \frac{\omega }{\omegaPO}
= \\
&=\frac{1}{r^2\omegaPO}\partial _r\left(
\visco r^3
\omegaPO
\partial _r\psi \right) +
\frac{1}{\omegaPO}
\partial _z\left(\visco
r\omegaPO
\partial _z\psi \right)
+ \frac{F_{\phi }}{\omegaPO}
\, .
\end{split}
\end{equation}
By virtue of the magnetic field form \reff{vectorb}, of the Lorentz force \reff{fphi},
the azimuthal equation stands at $r_0$ as 
\begin{equation}
\begin{split}
\epsilon r_0^2\left( v_rB_z - v_zB_r\right) +
2\epsilon v_r \frac{\omega _\mathrm{K}}{\omega ^{\prime }_0}
= r_0 \visco_{0}
\Delta \psi_1
+\\
+ \frac{1}{4\pi r_0^2\omega ^{\prime }_0}
\left[\partial _zI\left( \partial _{r_0}\psi_0
+ \partial _r\psi _1\right) 
 - \partial_rI\partial _z\psi_1\right]
\, ,
\label{exazeqlo}
\end{split}
\end{equation}
where $\visco_{0} =\visco(r_{0})$ and we neglected the $z$-derivatives of the viscosity coefficient in view of the hierarchy of the equations \cite{2010GReGr.tmp..112M}.
This relation accounts for the angular momentum transport
across the disk, by relying on the presence of a
viscous feature of the differential rotation.
Furthermore, this equation establishes a link between the radial and vertical
velocity fields with the poloidal currents present in the
configuration. 

A complete analysis of the dynamics is, however, still not possible. The system given by Eqs.\reff{verticalequilibrium}, \reff{radialequilibrium} and \reff{exazeqlo} is not closed, and we have to impose some other condition in order to describe the poloidal component of the magnetic field $I$ and the function $\Theta$. This task will be accomplished in the next Section when the balance of the forces acting on the electrons will be discussed.

\section{The electron force balance equation}\label{electronforcebalance}

In the presence of a non-zero resistivity coefficient
$\eta$, the equation accounting for the electron force
balance reads as
\begin{equation}
\vec{E} + \frac{\vec{v}}{c}\wedge \vec{B} =
\eta \vec{J}
\, . 
\label{efb1}
\end{equation}
Since the contribution of the resistive term is expected to be relatively small,
it turns out as appreciable only in the azimuthal component
of the equation above.
Thus, the balance of the Lorentz force has 
dominant radial and vertical components, 
providing the electric field in the form
predicted by the corotation theorem, i.e.
\begin{equation}
\vec{E} = -\frac{\vec{v}}{c}\wedge \vec{B} =
-\frac{\omega }{c}\left( \partial _r\psi \vec{e}_r
+ \partial _z\psi \vec{e}_z\right)
\, .
\label{MHDeqcond}
\end{equation}
Since the axial symmetry requires
$E_{\phi}\equiv 0$,
the $\phi$-component of the equation above
takes the form 
\begin{equation}
v_zB_r - v_rB_z = c\eta J_{\phi }
\, ,
\label{efb1x}
\end{equation}
and, taking into account \eref{jphi}, it reduces to 
\begin{equation}
v_rB_z - v_zB_r = 
\frac{\eta c^2}{4\pi r_0}
\Delta\psi_1
\, .
\label{efbloc}
\end{equation}
This relation, together with the azimuthal equilibrium equation (\ref{exazeqlo}), provides a system for the two unknowns
$\Theta$ and $I$, when $\psi_1$ and $\epsilon$
are given by the vertical \reff{verticalequilibrium} and radial \reff{radialequilibrium} equilibria. 
More precisely, 
substituting \eref{efbloc} in \eref{exazeqlo}, we get 
\begin{equation}
\begin{split}
2\epsilon v_r \frac{\omega _\mathrm{K}}{\omega ^{\prime }_0}
= r_0\left( \visco_{0} - \frac{c^2\eta \epsilon}{4\pi}\right) 
\Delta\psi_1 +
\\
+ \frac{1}{4\pi r_0^2\omega ^{\prime }_0}
\left[\partial _zI\left( \partial _{r_0}\psi_0
+ \partial _r\psi _1\right) 
 - \partial_rI\partial _z\psi_1\right]
\, .
\label{exazeqlocomb}
\end{split}\end{equation}


\section{Non-linear Configuration\label{sec:nonlinearapprox}}

In order to analyze the dynamics, we need to state some relation between the viscosity $\visco$ and the resistivity  $\eta$ of the plasma. As already done in  \cite{2010GReGr.tmp..112M}, we assume that the magnetic Prandtl number of the plasma is of order one. This in turns implies that we can take the following relation
\begin{equation}
\eta = 
\frac{4\pi \visco_{0}}{c^2(\bar{\epsilon}
+ \hat{\epsilon})}
\, .
\label{fundcond}
\end{equation}

Now, inserting \eref{fundcond} into \eref{exazeqlocomb},  we can get a direct relation between the radial velocity $v_r$, that accounts for the accretion on the central object, and the Lorentz force $F_{\phi}$; more precisely, we get
\begin{equation}
v_r = \frac{1}{2\omegaK(\bar{\epsilon}
+ \hat{\epsilon})}  F_{\phi }
\, .
\label{exazeqlocombn}
\end{equation}
Substituting $v_r$ by this equation into the
electron force balance \reff{efbloc}, we get the following
expression for $v_z$
\begin{equation}
\begin{split}
v_z = \frac{1}{\partial _z\psi _1(\bar{\epsilon}
+ \hat{\epsilon})}     
\left[\visco_{0}
\Delta\psi_1 - 
\left( \partial _{r_0}\psi _0 + \partial _r\psi _1\right) 
\frac{F_{\phi}}{2\omegaK}\right]
.
\end{split}
\label{efblocn}
\end{equation}
In terms of the function $\Theta$, the expressions
above for $v_r$ (\ref{exazeqlocombn}) and $v_z$ (\ref{efblocn}) stands as
\begin{eqnarray}
\label{thetaformofv}
&\partial _z\Theta = -\displaystyle\frac{r_0}{2\omega _\mathrm{K}} 
F_{\phi }\,,\\
&\partial _r\Theta = \displaystyle\frac{r_0}{\partial _z\psi _1}
\left[ \visco_{0}
\Delta\psi_1
- \frac{1}{2\omega _\mathrm{K}}
\left( \partial _{r_0}\psi _0 + \partial _r\psi _1\right) 
F_{\phi } \right]
 .
\end{eqnarray}
The solution of the first equation can be taken as
\begin{equation}
\Theta = -\frac{r_0}{2\omega _\mathrm{K}}
\int d z(F_{\phi })
\, ,
\label{thetafuncnl}
\end{equation}
which, substituted in the second, yields an
integro-differential relation for $F_{\phi }$, i.e.
\begin{equation}
\begin{split}
\partial _z\psi _1
\int d z(\partial _rF_{\phi }) = \left( \partial _{r_0}\psi _0 + \partial _r\psi _1\right) 
F_{\phi }
-2\omega _\mathrm{K}\visco_{0}\Delta\psi_1 \, .
\end{split}
\label{fphicomp}
\end{equation}
Once the behavior of $\psi _1$ is provided, from the equation above we can determine
the form of $F_{\phi }(r,\, z)$ and eventually
calculate the $\phi$-component $I(\psi ,\, z)$ of the magnetic field.

\subsection{Dimensionless variables}

Let us define the dimensionless functions 
$Y$, $\hat{D}$ and $\hat{P}$, in place of
$\psi _1$, $\hat{\epsilon}$ and $\hat{p}$, i.e.
\begin{equation}
Y\equiv \frac{k_0\psi _1}{\partial _{r_0}\psi _0}
\, , \hspace{5mm}
\hat{D}\equiv\beta \frac{ \hat{\epsilon}}{\epsilon _0}
\, , \hspace{5mm}
\hat{P}\equiv \beta \frac{\hat{p}}{p_0}
\, ,
\label{deff}
\end{equation}
where 
\begin{subequations}
\begin{eqnarray}
&\beta \equiv 8\pi \displaystyle\frac{p_0}{B^2_{0z}} = \displaystyle\frac{1}{3\epsilon _z^2} \equiv k_0^2\displaystyle\frac{H^2}{3}\,,&\\
&p_0\equiv 2\Kb\bar{T}\epsilon _0\,,\quad k_0\equiv 3\displaystyle\frac{\omegaK^2}{\vA^2}\,,\quad\vA^2\equiv \displaystyle\frac{B^2_{z0}}{4\pi \epsilon _0}\,,&
\end{eqnarray}
\end{subequations}
and  $B_{z0} = B_{z}(r, z=0) = \partial _{r_0}\psi _0/{r_0}$. We introduced the fundamental wavenumber $k_0$ that quantifies the typical length scale of the perturbation along the radial directions.
Finally, we rescale the
radial variable $x\equiv k_0(r - r_0)$, and the vertical one $u\equiv z/L_z$ assuming  the fundamental length in the vertical direction to be $L_z \equiv \sqrt{\epsilon _z}H$.

By these definitions, the vertical \reff{verticalequilibrium} and the radial equilibria \reff{radialequilibrium}
can be restated as
\begin{subequations}\label{sistema1}
\begin{equation}
\partial _{u^2}\hat{P} + \epsilon _z\hat{D}
+ 2\DeltaE Y
\partial _{u^2}Y = 0
\, ,
\label{vertad}
\end{equation}
\begin{equation}
\left(D + \frac{1}{\beta }\hat{D}\right) Y + 
\DeltaE Y (1 + \partial_xY )
+\frac{1}{2}\partial _x\hat{P} 
= 0 
\, ,
\label{radad}
\end{equation}
\end{subequations}
where 
\begin{equation}
\DeltaE Y\equiv\partial ^2_{x}Y + \epsilon _z\partial ^2_{u}Y\,.
\end{equation}
The two equations above provide a coupled system for $\hat{P}$ and $Y$ once the quantities $D$ and $\hat{D}$ are assigned;  hence we are able to determine the disk
configuration induced by the toroidal currents.
Finally, \eref{fphicomp} takes the form
\begin{equation}
\partial _uY
\int du(\partial _xA_{\phi }) =
-\DeltaE Y+ \left( 1 + \partial _xY\right) A_{\phi }
\, , 
\label{fphicompdim}
\end{equation}
$A_{\phi }\equiv F_{\phi }/2\omega _\mathrm{K}\visco_{0}k_0$
being a dimensionless function.

\section{The ring sequence}

\begin{figure}
\centering
\includegraphics[width=.95\columnwidth]{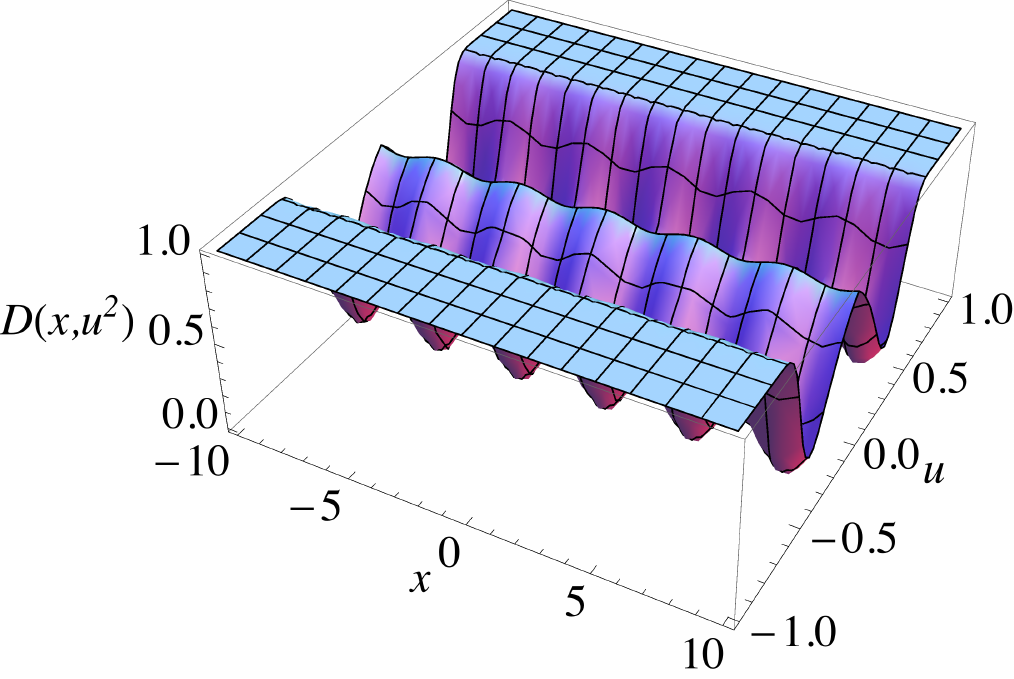}
\caption{3D-plot  for the total density $D\<{x,u^2}$ \reff{JacopoDFinale}, with parameters chosen as $A = 1\,, C = -1\,, a = -\sqrt{3}\,,b = -\tfrac{1}{6}\,,\varepsilon_z = 0.05\,, B = \sqrt{40}$.  The density profile exhibits an modulation along the radial direction, while in the vertical direction two minima appear.}
\label{fig:JacopoDFinale}
\end{figure}

\begin{figure}
\centering
\includegraphics[width=.8\columnwidth]{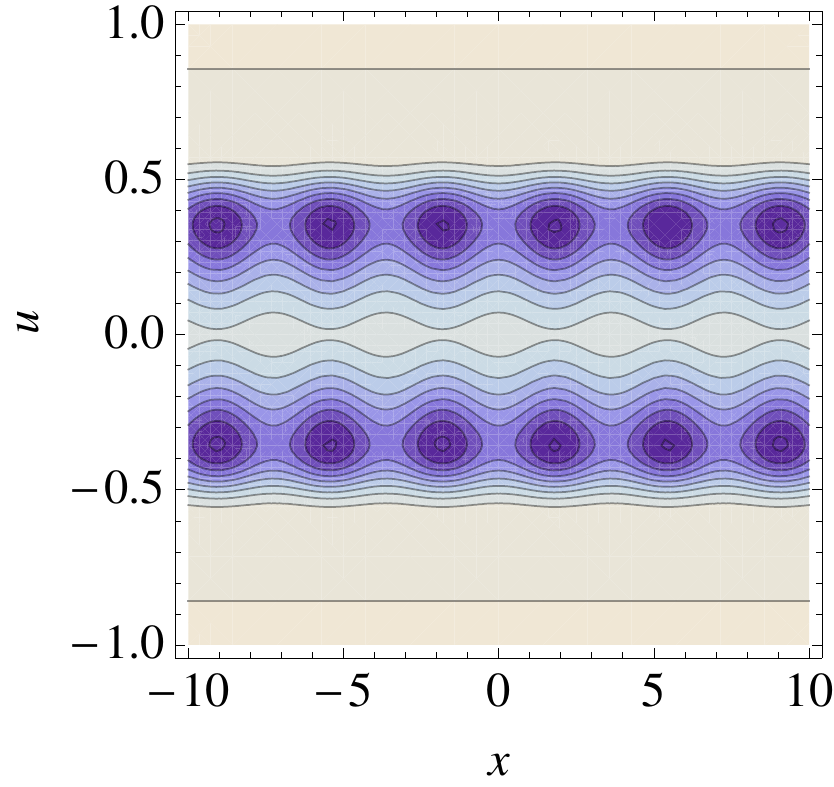}
\caption{Level lines for the total density $D\<{x,u^2}$ \reff{JacopoDFinale}, with parameters chosen as in \fref{fig:JacopoDFinale}. We note that the darkest zones are the minima of density, and the ring sequence appears in the equatorial plane.}
\label{fig:JacopoDFinale2}
\end{figure}

Here we  will derive a solution  to the system \reff{sistema1} in the case $\epsilon_z \neq 0$, showing how the formation of the crystalline structure for the magnetic field and the decomposition of the disk in a ring sequence can happen also in different regimes from those addressed in \cite{C05,CR06,2010GReGr.tmp..112M}.
Let us look for a particular solution of the form
\begin{subequations}\label{Jacopo2}
\begin{equation}\label{Jacopo2a}
Y\<{x,u^2}=F\<{u^2} \sin\<{a x}\,,
\end{equation}
\begin{equation}\label{Jacopo2b}
\hat{D}\<{x,u^2}=\Gamma\<{u^2} e^{ b \cos\<{a x}}\,,
\end{equation}
\begin{equation}\label{Jacopo2c}
\hat{P}\<{x,u^2}=M\<{u^2} \sin^2\<{a x} + N\<{u^2} e^{ b \cos \<{a x}} \, ,
\end{equation}
\end{subequations}
where $a, b$ are some parameters to be fixed.
We assume to be very close to the equatorial plane, in order to approximate  the function $\bar{D}$ to a constant
\begin{equation}\label{JacopoDensita}
\abs{u} \ll \frac{1}{\sqrt{\varepsilon_z}} \quad \Rightarrow  \quad \bar{D}\<{u^2}=\exp\<{-\varepsilon_z u^2} \simeq 1 \, .
\end{equation}
Inserting \erefs{Jacopo2} in \reff{sistema1},  the following system can be easily obtained
\begin{subequations}\label{Jacopo4}
\begin{equation} \label{Jacopo4a}
F'' - B^2 F = 0\,,
\qquad
M=F^2\,,
\end{equation}
\begin{equation}\label{Jacopo4c}
\frac{6 \varepsilon_z F}{a b} \dot{N} + N = 0\,,
\quad
\Gamma= -\frac{\dot{N}}{\varepsilon_z} \, ,
\end{equation}
\end{subequations}
where $()'\equiv d/d u$,  $\dot{()}\equiv d/d u^2$, and  $B$  a positive constant
\begin{equation} \label{costante_B}
B^2= \frac{a^2 - 1}{\varepsilon_z} > 0 \qquad \Rightarrow \qquad a^2 > 1 \, .
\end{equation}
The solution to system \reff{Jacopo4}  reads as
\begin{subequations}\label{soluzioniJacopo}
\begin{equation}
F=A e^{-B \sqrt{u^2}}\,,
\end{equation}
\begin{equation}
M=A^2 e^{-2B\sqrt{u^2}}\,,
\end{equation}
\begin{equation}
N = C \exp\<{ \frac{a b}{3 A \left(a^2 - 1\right)} e^{B\sqrt{u^2}} \left(1- B \sqrt{u^2}\right) }\,,
\end{equation}
\begin{equation}
\Gamma = \frac{a b C}{ 6 \varepsilon_z^2 A} \exp\left[\frac{a b  \left(1 - B \sqrt{u^2}\right) e^{B\sqrt{u^2}}}{3 A \left(a^2 - 1\right)}  + B \sqrt{u^2} \right]\,,
\end{equation}
\end{subequations}
where $A$ and $C$ are  integration constants. 
The finiteness of the solutions, leads us to require
\begin{equation} \label{no_divergenza}
\frac{a b}{A} > 0 \, .
\end{equation}
Finally, the total mass density assumes the following form
\begin{equation}\label{JacopoDFinale}
\begin{split}
D&\<{x,u^2}=\bar{D} + \displaystyle\frac{1}{\beta} \hat{D} = \\
&=1+  \frac{a b C}{ 2 A} \exp\left[-\frac{a b e^{B\sqrt{u^2}}}{3 A \left(a^2 - 1\right)}  \left( B \sqrt{u^2} - 1\right)\right]\times \\
&\times\exp\<{B \sqrt{u^2} + b \cos\<{a x} } \, ,
\end{split}
\end{equation}
which results to be positive definite if the following condition holds:
\begin{equation} \label{d_positiva}
\abs{\frac{a b C}{ 2 A}} \exp\<{ - \frac{a b}{3 A \left(a^2 - 1\right)} + \abs{b}} < 1 \, .
\end{equation}
The formation of the ring sequence can be seen in Figs.\reff{fig:JacopoDFinale} and \reff{fig:JacopoDFinale2} representing the density \reff{JacopoDFinale} for a particular choice of the model parameters.

It is worth noting that solutions \reff{soluzioniJacopo} satisfy also \eref{fphicompdim} as soon as we fix the form of the Lorentz force as the following
\begin{equation}\label{ultima}
A_\phi \<{x,u^2} = - F\<{u^2} \sin \<{a x} \, .
\end{equation}
Finally, from \eref{exazeqlocombn}, we  obtain a non-zero radial velocity
\begin{equation} \label{JacopoVR}
v_r = - \frac{A \visco_0 k_0}{\epsilon_0} \frac{e^{-B \sqrt{u^2}} \sin \<{a x}}{D} \,.
\end{equation}

\section{The corotation theorem for a weakly resistive and viscous plasma}\label{sec:dim}

In the following,  we  will show how the corotation theorem by Ferraro \cite{F37} holds under the hypothesis of stationarity up to the second order in  small dissipative effects, poloidal velocity and toroidal component of the magnetic field.

Let us consider the Maxwell equations
\begin{eqnarray}\label{dimferraro1}
&\eta \vec{J} = \vec{E} + \displaystyle\frac{1}{c} \vec{v} \wedge \vec{B}\,,& \\
&\displaystyle\grad\wedge\vec{B} = \frac{4 \pi}{c} \vec{J}\,,&
\end{eqnarray}
and combine them by eliminating $\vJ$
\begin{equation}\label{irrotational}
c \vec{E} = \frac{c^2}{4 \pi} \eta \grad\wedge\vec{B} - \vec{v} \wedge \vB \, .
\end{equation}
In the case of a steady-state configuration, the vector $\vec{E}$ is irrotational. So, as soon as the poloidal velocity and the toroidal magnetic component are small enough, and by using \eref{solconteq} and \eref{vectorb}, \eref{irrotational} reduces to 
\begin{equation}\label{dimferraro3}
\grad \wedge
\left(
\begin{array}{c}
\displaystyle\frac{4 \pi}{c^2} \omega \partial_r \psi + \eta \frac{1}{r} \partial_z I \\
\\
\displaystyle\frac{4 \pi}{c^2} \left( v_r B_z - v_z B_r \right) + \eta \left[\frac{\partial_z^2 \psi}{r}  + \partial_r \left(\frac{\partial_r \psi}{r}\right)\right]\\
\\
\displaystyle\frac{4 \pi}{c^2} \omega \partial_z \psi - \eta \frac{1}{r} \partial_r I
\end{array}
\right)
= 0\,.
\end{equation}
The radial and vertical components of \eref{dimferraro3}, which are of first order in the small quantities, are identically satisfied as soon as one imposes the electron force balance  \reff{efb1x}. On the other hand, the azimuthal equation can be rewritten for convenience in the vector form
\begin{equation}\label{dimferraro5}
\frac{4 \pi}{c^2} \left( \grad \omega \wedge \grad \psi \right) = \grad \wedge \left( \eta \frac{ \grad I}{r} \wedge \hat{\phi} \right)\,.
\end{equation}

The corotation theorem, stating that $\grad \omega \wedge \grad \psi = 0$, can be recovered up to the first order because the right hand side of the last equation is of second order in $\eta$, $I$. Indeed, we have to require not only $\eta$ and $I$ to be small, but also that their gradients must retain the same behavior. It is possible to check that this is the case allowed in the solution derived above  (see the crossmath of \eref{fphi} and \eref{ultima}).

\section{Concluding Remarks}

We analyze an axis-symmetric thin disk configuration whose structure is summarized by a mainly rotating plasma embedded in a magnetic field endowed with a small toroidal component. The consistence of the configuration scheme requires that correspondingly small poloidal matter fluxes and dissipative effects must be included in the problem. This analysis, which generalizes the original approach by Coppi and collaborators, has the merit of providing a coherent description of the vertical and radial dependence of the plasma profile.
Matter fluxes result to be necessary when speaking of a real accreting disk, while the toroidal component of the magnetic field is expected to be non zero when dealing with the magnetic configuration of an astrophysical source. The profile we obtained corresponds to have a magnetic Prandtl number of order unity (which is a typical value required to phenomenologically address the observations \cite{Sp08}), and it offers the natural scenario where a perturbation scheme of the ring sequence profile can be implemented. 

This work was developed within the framework of the CGW Collaboration (http:// www.cgwcollaboration.it).


\newcommand{\cqg}{Class. Quant. Grav}
\newcommand{\prl}{Phys. Rev. Lett.}
\newcommand{\pra}{Phys. Rev. A}
\newcommand{\prb}{Phys. Rev. B}
\newcommand{\prc}{Phys. Rev. C}
\newcommand{\prd}{Phys. Rev. D}
\newcommand{\pre}{Phys. Rev. E}
\newcommand{\PR}{Phys. Rev.}
\newcommand{\apj}{Astrophys. J.}
\newcommand{\apjl}{Astrophys. J. Lett.}
\newcommand{\araa}{Ann. Rev. A.\& A.}
\newcommand{\aj}{Astron. J.}
\newcommand{\apss}{Astrophys.  Space Science}
\newcommand{\physrep}{Phys. Rep.}
\newcommand{\mnras}{MNRAS}
\newcommand{\azh}{Astronomicheskii Zhurnal}
\newcommand{\aap}{A.\&A.}
\newcommand{\nat}{Nat.}
\newcommand{\ptp}{Progr. Theor. Phys.}
\newcommand{\ijtp}{Int. J.  Theor. Phys.}
\newcommand{\grg}{Gen. Rel. Grav.}
\newcommand{\pla}{Phys. Lett. A}
\newcommand{\plb}{Phys. Lett. B}
\newcommand{\plc}{Phys. Lett. C}
\newcommand{\pld}{Phys. Lett. D}
\newcommand{\ple}{Phys. Lett. E}
\newcommand{\epl}{Europhys. Lett.}
\newcommand{\ijmpa}{Int. J. Mod. Phys. A}
\newcommand{\ijmpb}{Int. J. Mod. Phys. B}
\newcommand{\ijmpc}{Int. J. Mod. Phys. C}
\newcommand{\ijmpd}{Int. J. Mod. Phys. D}
\newcommand{\ijmpe}{Int. J. Mod. Phys. E}
\newcommand{\mpla}{Mod. Phys. Lett. A}
\newcommand{\mplb}{Mod. Phys. Lett. B}
\newcommand{\advphys}{Adv. Phys.}
\newcommand{\sovJETP}{Sov. Phy. JETP}
\newcommand{\npb}{Nucl. Phys. B}
\newcommand{\JETPl}{JETP Lett.}
\newcommand{\zetf}{Zhurn. Eks. Teor. Fiz.}
\newcommand{\pzetf}{Pis ma Zhurn. Eks. Teor. Fiz.}
\newcommand{\sjetp}{Sov. J. Exp. Theor. Phys.}
\newcommand{\bain}{Bull. Astron. Inst. Netherl.}
\newcommand{\jpa}{J. Phys. A}
\newcommand{\jpb}{J. Phys. B}
\newcommand{\jhep}{J. High En. Phys.}
\newcommand{\lrr}{Liv. Rev. Rel.}
\newcommand{\rpp}{Rep. Progr. Phys.}
\newcommand{\lmp}{Lett. Math. Phys.}
\newcommand{\atmp}{Adv. Theor. Math. Phys.}
\newcommand{\PoS}{Proc. Science}
\newcommand{\rmp}{Rev. Mod. Phys.}
\newcommand{\ahp}{Ann. Henri Poincar\'e}
\newcommand{\GaC}{Grav. Cosm.}
\newcommand{\asr}{Adv. Space Res.}
\newcommand{\ppcf}{Plasma Phys. Contr. Fus.}
\newcommand{\PP}{Phys. Plasmas}
\newcommand{\SA}{Sov. Astron.}
\newcommand{\nar}{New Astron. Rev.}
\newcommand{\NF}{Nucl. Fus.}
\newcommand{\ssr}{Space Science Rev.}


\begin{thebibliography}{23}
\expandafter\ifx\csname natexlab\endcsname\relax\def\natexlab#1{#1}\fi
\expandafter\ifx\csname bibnamefont\endcsname\relax
  \def\bibnamefont#1{#1}\fi
\expandafter\ifx\csname bibfnamefont\endcsname\relax
  \def\bibfnamefont#1{#1}\fi
\expandafter\ifx\csname citenamefont\endcsname\relax
  \def\citenamefont#1{#1}\fi
\expandafter\ifx\csname url\endcsname\relax
  \def\url#1{\texttt{#1}}\fi
\expandafter\ifx\csname urlprefix\endcsname\relax\def\urlprefix{URL }\fi
\providecommand{\bibinfo}[2]{#2}
\providecommand{\eprint}[2][]{\url{#2}}

\bibitem[{\citenamefont{{Balbus} and {Hawley}}(1991)}]{1991ApJ...376..214B}
\bibinfo{author}{\bibnamefont{{Balbus}}, \bibfnamefont{S.~A.}}, and
  \bibinfo{author}{\bibfnamefont{J.~F.} \bibnamefont{{Hawley}}},
  \bibinfo{year}{1991}, \bibinfo{journal}{\apj} \textbf{\bibinfo{volume}{376}},
  \bibinfo{pages}{214}.

\bibitem[{\citenamefont{{Balbus} and {Hawley}}(1998)}]{B98}
\bibinfo{author}{\bibnamefont{{Balbus}}, \bibfnamefont{S.~A.}}, and
  \bibinfo{author}{\bibfnamefont{J.~F.} \bibnamefont{{Hawley}}},
  \bibinfo{year}{1998}, \bibinfo{journal}{\rmp} \textbf{\bibinfo{volume}{70}},
  \bibinfo{pages}{1}.

\bibitem[{\citenamefont{{Bisnovatyi-Kogan} and
  {Lovelace}}(2000)}]{2000ApJ...529..978B}
\bibinfo{author}{\bibnamefont{{Bisnovatyi-Kogan}}, \bibfnamefont{G.~S.}}, and
  \bibinfo{author}{\bibfnamefont{R.~V.~E.} \bibnamefont{{Lovelace}}},
  \bibinfo{year}{2000}, \bibinfo{journal}{\apj} \textbf{\bibinfo{volume}{529}},
  \bibinfo{pages}{978}.

\bibitem[{\citenamefont{{Bisnovatyi-Kogan} and {Lovelace}}(2001)}]{Bisno01}
\bibinfo{author}{\bibnamefont{{Bisnovatyi-Kogan}}, \bibfnamefont{G.~S.}}, and
  \bibinfo{author}{\bibfnamefont{R.~V.~E.} \bibnamefont{{Lovelace}}},
  \bibinfo{year}{2001}, \bibinfo{journal}{\nar} \textbf{\bibinfo{volume}{45}},
  \bibinfo{pages}{663}.

\bibitem[{\citenamefont{{Chandrasekhar}}(1960)}]{1960PNAS...46..253C}
\bibinfo{author}{\bibnamefont{{Chandrasekhar}}, \bibfnamefont{S.}},
  \bibinfo{year}{1960}, \bibinfo{journal}{Proceedings of the National Academy
  of Science} \textbf{\bibinfo{volume}{46}}, \bibinfo{pages}{253}.

\bibitem[{\citenamefont{{Coppi}}(2005)}]{C05}
\bibinfo{author}{\bibnamefont{{Coppi}}, \bibfnamefont{B.}},
  \bibinfo{year}{2005}, \bibinfo{journal}{\PP} \textbf{\bibinfo{volume}{12}},
  \bibinfo{pages}{7302}.

\bibitem[{\citenamefont{{Coppi}}(2009)}]{CoppiHighEnergy}
\bibinfo{author}{\bibnamefont{{Coppi}}, \bibfnamefont{B.}},
  \bibinfo{year}{2009}, \bibinfo{journal}{\ppcf}
  \textbf{\bibinfo{volume}{51}}(\bibinfo{number}{12}), \bibinfo{pages}{124007}.

\bibitem[{\citenamefont{{Coppi} and {Coppi}}(2001)}]{Coppi:2001p102}
\bibinfo{author}{\bibnamefont{{Coppi}}, \bibfnamefont{B.}}, and
  \bibinfo{author}{\bibfnamefont{P.~S.} \bibnamefont{{Coppi}}},
  \bibinfo{year}{2001}, \bibinfo{journal}{\prl}
  \textbf{\bibinfo{volume}{87}}(\bibinfo{number}{5}), \bibinfo{pages}{051101}.

\bibitem[{\citenamefont{{Coppi} and Keyes}(2003)}]{Coppi:2003p112}
\bibinfo{author}{\bibnamefont{{Coppi}}, \bibfnamefont{B.}}, and
  \bibinfo{author}{\bibfnamefont{E.~A.} \bibnamefont{Keyes}},
  \bibinfo{year}{2003}, \bibinfo{journal}{\apj} \textbf{\bibinfo{volume}{595}},
  \bibinfo{pages}{1000}.

\bibitem[{\citenamefont{{Coppi} and {Rousseau}}(2006)}]{CR06}
\bibinfo{author}{\bibnamefont{{Coppi}}, \bibfnamefont{B.}}, and
  \bibinfo{author}{\bibfnamefont{F.}~\bibnamefont{{Rousseau}}},
  \bibinfo{year}{2006}, \bibinfo{journal}{\apj} \textbf{\bibinfo{volume}{641}},
  \bibinfo{pages}{458}.

\bibitem[{\citenamefont{{Ferraro}}(1937)}]{F37}
\bibinfo{author}{\bibnamefont{{Ferraro}}, \bibfnamefont{V.~C.~A.}},
  \bibinfo{year}{1937}, \bibinfo{journal}{\mnras}
  \textbf{\bibinfo{volume}{97}}, \bibinfo{pages}{458}.

\bibitem[{\citenamefont{{Krolik}}(1999)}]{LibroAGN}
\bibinfo{author}{\bibnamefont{{Krolik}}, \bibfnamefont{J.~H.}},
  \bibinfo{year}{1999}, \emph{\bibinfo{title}{{Active galactic nuclei: from the
  central black hole to the galactic environment}}}
  (\bibinfo{publisher}{Princeton Univ. Press}).

\bibitem[{\citenamefont{{Lynden-Bell}}(1996)}]{L96}
\bibinfo{author}{\bibnamefont{{Lynden-Bell}}, \bibfnamefont{D.}},
  \bibinfo{year}{1996}, \bibinfo{journal}{\mnras}
  \textbf{\bibinfo{volume}{279}}, \bibinfo{pages}{389}.

\bibitem[{\citenamefont{{Lynden-Bell} and {Pringle}}(1974)}]{lyndenpringle74}
\bibinfo{author}{\bibnamefont{{Lynden-Bell}}, \bibfnamefont{D.}}, and
  \bibinfo{author}{\bibfnamefont{J.~E.} \bibnamefont{{Pringle}}},
  \bibinfo{year}{1974}, \bibinfo{journal}{\mnras}
  \textbf{\bibinfo{volume}{168}}, \bibinfo{pages}{603}.

\bibitem[{\citenamefont{{Montani} and {Benini}}(2011)}]{2011PhRvE..84b6406M}
\bibinfo{author}{\bibnamefont{{Montani}}, \bibfnamefont{G.}}, and
  \bibinfo{author}{\bibfnamefont{R.}~\bibnamefont{{Benini}}},
  \bibinfo{year}{2011}, \bibinfo{journal}{\pre}
  \textbf{\bibinfo{volume}{84}}(\bibinfo{number}{2}), \bibinfo{pages}{026406}.

\bibitem[{\citenamefont{Montani and Benini}(2011)}]{2010GReGr.tmp..112M}
\bibinfo{author}{\bibnamefont{Montani}, \bibfnamefont{G.}}, and
  \bibinfo{author}{\bibfnamefont{R.}~\bibnamefont{Benini}},
  \bibinfo{year}{2011}, \bibinfo{journal}{\grg} \textbf{\bibinfo{volume}{43}},
  \bibinfo{pages}{1121}.

\bibitem[{\citenamefont{{Montani} and {Carlevaro}}(2010)}]{2010PhRvE..82b5402M}
\bibinfo{author}{\bibnamefont{{Montani}}, \bibfnamefont{G.}}, and
  \bibinfo{author}{\bibfnamefont{N.}~\bibnamefont{{Carlevaro}}},
  \bibinfo{year}{2010}, \bibinfo{journal}{\pre}
  \textbf{\bibinfo{volume}{82}} (\bibinfo{number}{2}), \bibinfo{pages}{025402}.

\bibitem[{\citenamefont{{Piran}}(1999)}]{ReviewPiran}
\bibinfo{author}{\bibnamefont{{Piran}}, \bibfnamefont{T.}},
  \bibinfo{year}{1999}, \bibinfo{journal}{\physrep}
  \textbf{\bibinfo{volume}{314}}, \bibinfo{pages}{575}.

\bibitem[{\citenamefont{Pringle and Rees}(1972)}]{pringlerees72}
\bibinfo{author}{\bibnamefont{Pringle}, \bibfnamefont{J.~E.}}, and
  \bibinfo{author}{\bibfnamefont{M.~J.} \bibnamefont{Rees}},
  \bibinfo{year}{1972}, \bibinfo{journal}{\aap} \textbf{\bibinfo{volume}{21}},
  \bibinfo{pages}{1}.

\bibitem[{\citenamefont{Shakura}(1973)}]{S73}
\bibinfo{author}{\bibnamefont{Shakura}, \bibfnamefont{N.~I.}},
  \bibinfo{year}{1973}, \bibinfo{journal}{\SA} \textbf{\bibinfo{volume}{16}},
  \bibinfo{pages}{756}.

\bibitem[{\citenamefont{{Shakura} and {Sunyaev}}(1988)}]{Shakura:1988p104}
\bibinfo{author}{\bibnamefont{{Shakura}}, \bibfnamefont{N.~I.}}, and
  \bibinfo{author}{\bibfnamefont{R.~A.} \bibnamefont{{Sunyaev}}},
  \bibinfo{year}{1988}, \bibinfo{journal}{\asr}
  \textbf{\bibinfo{volume}{8}}(\bibinfo{number}{2-3}), \bibinfo{pages}{135}.

\bibitem[{\citenamefont{Spruit}(2010)}]{Sp08}
\bibinfo{author}{\bibnamefont{Spruit}, \bibfnamefont{H.}},
  \bibinfo{year}{2010}, \bibinfo{journal}{Lect.Notes Phys.}
  \textbf{\bibinfo{volume}{794}}, \bibinfo{pages}{233}.

\bibitem[{\citenamefont{Velikhov}(1959)}]{V59}
\bibinfo{author}{\bibnamefont{Velikhov}, \bibfnamefont{E.}},
  \bibinfo{year}{1959}, \bibinfo{journal}{Sov. Phys. JETP}
  \textbf{\bibinfo{volume}{36}}(\bibinfo{number}{9}), \bibinfo{pages}{995}.

\end{thebibliography}
\end{document}